\documentclass[twocolumn]{jpsj2}
\usepackage{graphicx}
\usepackage[usenames]{color}
\newcommand{\bfk}{{\mib k}}
\newcommand{\bfq}{{\mib q}}

\newcommand{\bfr}{{\mib r}}

\def\grtsim{\raisebox{0.5ex}{$>$}\hspace{-3.0mm}\raisebox{-0.5ex}{$\sim$}}

\title
{
Nonmagnetic-Defect-Induced Magnetism in Graphene
}

\author
{ 
Hideki {\sc Kumazaki} and 
Dai S. {\sc Hirashima}
}

\inst
{
Department of Physics, Nagoya University, Nagoya 464-8602.
}

\recdate
{
\today
}

\abst
{
It is shown that a strong impurity potential induces short-range
antiferromagnetic (ferrimagnetic) order around itself in a Hubbard
model on a half-filled honeycomb lattice. 
This implies that short-range magnetic order
is induced in monolayer graphene 
by a nonmagnetic defect such as a vacancy with full hydrogen termination
or a chemisorption defect.
}

\kword
{
graphene, vacancies, magnetic moment
}
\begin{document}
\sloppy
\maketitle
%
Physical properties of graphene (a monolayer graphite sheet) have
attracted much interest since it was fabricated,\cite{Nov0} 
magneto-transport
properties were studied,\cite{Nov,Zhang} and 
various peculiar properties such as the
half-integer quantum Hall effect\cite{Zheng,Gusynin1,Peres1} were found.
Although the main interest is in its transport properties at present,
magnetic properties are also worth studying.
Indeed, interest in magnetism in carbon-based materials has recently
surged because of not only its fundamental importance, but also
possibility of application in new technologies.\cite{carb}

Electrons in graphene can be described by a tight-binding model
on a honeycomb lattice. 
The purpose of this paper is 
to show that short-range antiferromagnetic (AF) 
(more precisely, ferrimagnetic) order is 
induced around an impurity introduced on a half-filled honeycomb lattice.

Magnetism in graphite systems has been studied in many 
experiments,\cite{Shibay,Esqu} although the issue has been controversial;
the effect of magnetic impurities cannot be completely discarded.
Apart from magnetic impurities, the defect-induced mechanism is the
most probable mechanism of magnetism in carbon-based materials. 
For example, a free dangling bond associated with
a vacancy will have a net magnetic moment, which may induce a magnetic
state. Even if a defect itself is nonmagnetic, it can cause spin
polarization of $\pi$ electrons
around it. It has been known that localized edge
states are generated at a zigzag edge and they can
lead to a magnetic state.\cite{Fuji,Nakata,Oka}
In previous studies,\cite{kuma1,kuma2} we showed that a strong impurity
potential in a half-filled honeycomb lattice induces a localized
state (zero mode) around it\cite{Waka2,Pereira,Wehling} and that 
it causes enhancement of staggered susceptibility.
We did not find any (short-range) order, however, because we did not
consider electron-electron interaction. In this study,
considering electron-electron interaction, we show that short-range
magnetic order is indeed 
induced around a strong-impurity site in a half-filled honeycomb
lattice.

We start from a tight-binding model on a honeycomb lattice.
We consider only the nearest neighbor transfer $t$,
and then the hamiltonian is given by
\begin{equation}
{\cal H}_{0} = -t \sum_{(\mib{n}A,\mib{n}'B), \sigma} [
c_{\mib{n}A\sigma}^{\dagger} c_{\mib{n}B\sigma} + {\rm h.c.}],
\end{equation}
where
$c_{\mib{n}\alpha\sigma}$ ($c_{\mib{n}\alpha\sigma}^{\dagger}$) 
is an annihilation (creation) operator of an electron with the spin $\sigma$
on sublattice $\alpha$ in the $\mib{n}$th unit cell, and $(\mib{n}A,\mib{n'}B)$
stands for the pairs of the nearest sites.
We denote the number of the unit cell by $N_{\rm u}=L^{2}$. 
The number of the lattice
points is $N_{\rm L}=2N_{\rm u}=2L^{2}$.
In addition to ${\cal H}_{0}$, we consider electron-electron interaction.
We consider the on-site Coulomb interaction $U$. Long-range Coulomb
interaction is poorly screened in graphene. However, because it must be 
the short-range
part of the Coulomb interaction that plays the most important role in
inducing magnetism, we consider only the on-site Coulomb interaction
and use the following interaction
hamiltonian ${\cal H}'$,
\begin{equation}
{\cal H}'=U\sum_{\mib{n},\alpha} \hat{n}_{\mib{n}\alpha\uparrow} 
                                 \hat{n}_{\mib{n}\alpha\downarrow},
\end{equation}
where
$\hat{n}_{\mib{n}\alpha\sigma}=c_{\mib{n}\alpha\sigma}^{\dagger}
                         c_{\mib{n}\alpha\sigma}$.
We thus consider a Hubbard model on the honeycomb lattice.

Furthermore, we introduce impurities. We consider a
short-range impurity potential. If an impurity is on sublattice $\alpha$
in the $\mib{n}$th unit cell, its effect is expressed by
\begin{equation}
{\cal V}(\mib{n},\alpha)=u \sum_{\sigma}c^{\dagger}_{\mib{n}\alpha\sigma}
                           c_{\mib{n}\alpha\sigma}.
\end{equation}
The limit of $u\rightarrow \infty$ represents a strong impurity 
such as a vacancy or a hydrogen chemisorption defect.\cite{move}
In that case, the total number of the lattice points is
$N_{\rm L}=2N-N_{\rm i}$, where $N_{\rm i}$ is the number of impurities.
We consider the half-filled case, {\it i.e.}, the case where
the number $N_{\rm e}$ of electrons is equal to $N_{\rm L}$,
in this study. Actually, we consider the case with $N_{\rm i}=2$.

We resort to a mean field approximation. The interaction term
is approximated as
\begin{equation}
{\cal H}'\simeq -U\sum_{\mib{n},\alpha} 
\langle \hat{n}_{\mib{n}\alpha\uparrow} \rangle
\langle \hat{n}_{\mib{n}\alpha\downarrow}\rangle
+U\sum_{\mib{n},\alpha,\sigma} \langle \hat{n}_{\mib{n}\alpha-\sigma}\rangle
\hat{n}_{\mib{n}\alpha\sigma}.
\end{equation}
The mean field hamiltonian ${\cal H}_{\rm mf}$ is then given by,
apart from a c-number term,
\begin{equation}
{\cal H}_{\rm mf}=
\sum_{\mib{n}\alpha\sigma}\epsilon_{\mib{n}\alpha\sigma}\hat{n}_{\mib{n}\alpha
\sigma}+{\cal H}_{0}+{\cal V},
\end{equation}
where $\epsilon_{\mib{n}\alpha\sigma}=U\langle
\hat{n}_{\mib{n}\alpha-\sigma} \rangle.$
By diagonalizing ${\cal H}_{\rm mf}$, we self-consistently determine
$2N_{\rm L}$ parameters, $\epsilon_{\mib{n}\alpha\sigma}$, or,
equivalently, $\langle \hat{n}_{\mib{n}\alpha\sigma}\rangle$, from which
we can readily calculate the electron density $n_{\mib{n}
\alpha}=\langle
\hat{n}_{\mib{n}\alpha\uparrow}+\hat{n}_{\mib{n}\alpha\downarrow}\rangle$
and the spin density $m_{\mib{n}\alpha}=\langle
\hat{n}_{\mib{n}\alpha\uparrow}-\hat{n}_{\mib{n}\alpha\downarrow}\rangle$
at each lattice point.

\begin{figure}[hbt]
\begin{center}
\includegraphics[width=5cm]{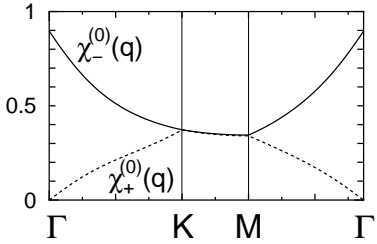}
\caption{Momentum dependence of spin susceptibilities $\chi_{+}(\bfq)$
(dotted curve) and $\chi_{-}(\bfq)$ (solid curve)
of the half-filled honeycomb lattice
at $U=0$ and $T=0$.
}
\label{fig:q-dep}
\end{center}
\end{figure}

Before showing the results in the presence of impurities, we briefly
study the impurity-free case.
We define static spin susceptibilities $\chi_{\pm}(\bfq)$ by
\begin{eqnarray}
\chi_{\pm}(\bfq)=\frac{1}{N_{\rm u}}
\int_{0}^{\beta}\!\!\!\!\!
&{\rm d}\tau&
\!\!\!\!\!
\langle
[\;\hat{m}_{A}(\bfq,\tau)\pm \hat{m}_{B}(\bfq,\tau)]
\nonumber \\
&\times& \!\!\!
[\hat{m}_{A}(-\bfq,0)\pm \hat{m}_{B}(-\bfq,0)]\;
\rangle,
\end{eqnarray}
where
\begin{equation}
\hat{m}_{\alpha}(\bfq)=\sum_{\bfk} (c_{\bfk-\bfq\alpha\uparrow}^{\dagger}
                             c_{\bfk-\bfq\alpha\uparrow}
-                            c_{\bfk-\bfq\alpha\downarrow}^{\dagger}
                             c_{\bfk-\bfq\alpha\downarrow}).
\end{equation}
$\chi_{+}(0)$ is uniform susceptibility and $\chi_{-}(0)$ is staggered
susceptibility.
In Fig. \ref{fig:q-dep}, we show the momentum dependence of 
$\chi_{\pm}(\bfq)$ for $U=0$ at absolute zero, $T=0$.
It can be seen that staggered susceptibility $\chi_{-}(\bfq=0)$ takes the
maximum value.

In the random phase approximation (RPA), spin susceptibilities can be
written as
\begin{equation}
\chi_{\pm}(\bfq)=\frac{\chi_{\pm}^{(0)}(\bfq)}
{1-\frac{U}{2}\chi_{\pm}^{(0)}(\bfq)},
\end{equation}
from which we can see that the paramagnetic state is unstable
towards an AF state at $U \ge U_{\rm cr}^{(0)}=
2/\chi_{-}^{(0)}(0)
\simeq 2.23 t$.

\begin{figure}[hbt]
\begin{center}
\includegraphics[width=5.5cm]{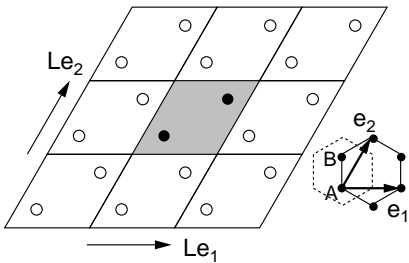}
\caption{
Honeycomb lattice of size $L\times L$ used in calculations.
The periodic boundary condition is imposed. Solid dots represents the
positions of the unit cells where impurities are placed, and open
dots are their images. $\mib{e}_{i}$ ($i=1,2$) is the unit vector of
length $a$, where $a$ is the lattice constant. The dotted line shows a
unit cell which includes an A sublattice point and a B sublattice point.
}
\label{fig:fig2}
\end{center}
\end{figure}
Now, we study the impurity effect using a finite size lattice.
The lattice with $L\times L$ unit cells is shown in Fig. \ref{fig:fig2}.
We place two impurities in the unit cells at $\mib{n}_{1}=(n_{1},n_{1})$ and
and $\mib{n}_{2}=\mib{n}_{1}+(L/2,L/2)$, and impose the periodic boundary
condition. As can be seen from Fig. \ref{fig:fig2}, the (shortest)
distance between the two impurity sites is $L/2$.

\begin{figure}[hbt]
\begin{center}
\includegraphics[width=4cm]{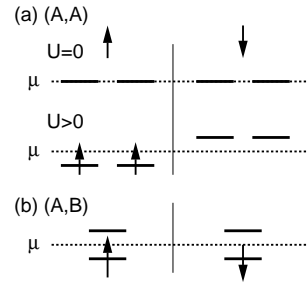}
\caption{The ground state spin configuration in the presence of two
vacancies (a) on one of the sublattices and (b) on sublattices A and B.
(a) At $U=0$, four zero modes
(two per each spin) appear at the chemical potential $\mu$. At $U>0$,
two of them are occupied by equal spins (up spins in this figure).
(b) Zero modes appear symmetrically around the chemical potential $\mu$,
and two of them are occupied by an up-spin and a down-spin.
}
\label{fig:AA_diagr}
\end{center}
\end{figure}

\begin{figure}[hbt]
\begin{center}
\includegraphics[width=7.5cm]{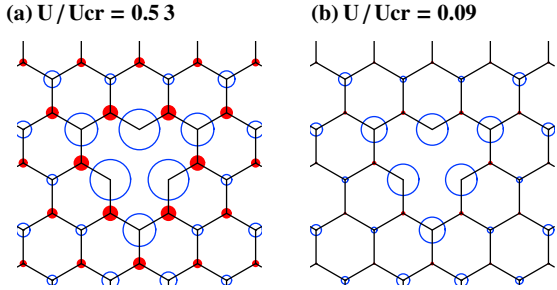}
\caption{Spin density around an impurity site at 
(a) $U=1.2 t=0.53 U_{\rm cr}^{(0)}$ and (b) $U=0.2 t=0.09 U_{\rm cr}^{(0)}$ for
$L=50$. Area of each dot is proportional to the magnitude of the
spin density at each lattice point.
Open dots represent positive magnetic moments and filled dots negative
magnetic moments.
}
\label{fig:spind}
\end{center}
\end{figure}

\begin{figure}[hbt]
\begin{center}
\includegraphics[width=5.5cm]{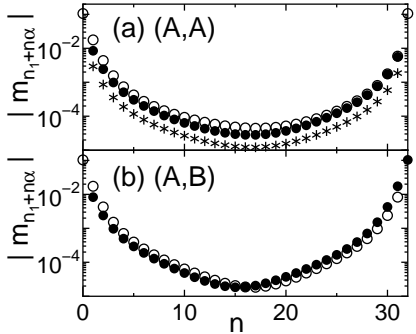}
\caption{
Moments $-m_{\mib{n}'A}$ (solid dots) and  $m_{\mib{n}'B}$ (open dots)
in the unit cells at $\mib{n}'=\mib{n}_{1}+(n,n)$ at $U=1.0 t$
for $L=64$. 
(a) Impurities are on the A sublattice points at $\mib{n}_{1}$ and at
$\mib{n}_{1}+(32,32)$. For comparison, $-m_{\mib{n}'A}$ at $U=0.5 t$
is also shown (stars).
(b) Impurities are on the A sublattice point at $\mib{n}_{1}$ and
on the B sublattice point at $\mib{n}_{1}+(32,32)$.
}
\label{fig:fig4}
\end{center}
\end{figure}

First, we study the case with
$u\rightarrow\infty$, {\it i.e.}, $N_{\rm L}=2N_{\rm u}-2$, and
deal with the case where both impurities are on the same
sublattice, say sublattice A; we denote this case by (A,A). 
In this case, the ground state
is fourfold degenerate at $U=0$, because four degenerate zero
modes $\phi_{0\sigma}^{(k)}(\bfr) (\sigma=\uparrow, \downarrow$
and $k=1,2$) appear just at the chemical 
potential $\mu$ (see Fig. \ref{fig:AA_diagr} (a)).
At a finite $U$, two of the zero modes are occupied by equal spins in the
ground state (Hund's rule), that is, 
the ground state always has a total
moment $M_{\rm t}=m_{A}+m_{B}=2$
where $m_{\alpha}$ is the total moment of the sublattice $\alpha$,
$m_{\alpha}=\sum_{\mib{n}}(n_{\mib{n}\alpha\uparrow}-n_{\mib{n}\alpha
\downarrow}$).
This conclusion is in harmony with Lieb's
theorem.\cite{Lieb}
Because the zero-mode wave function $\phi_{0\sigma}^{(k)}(\bfr)$ 
is localized
around impurity sites, the spin density is also localized around
impurity sites. In Fig. \ref{fig:spind}, 
we show the spin density around an impurity
site. It can be seen that short-range ferrimagnetic order
is induced around the impurity site and that the spin density on 
sublattice B is larger than that on sublattice A, the sublattice where
the impurity is.\cite{charge}
As a finite total moment $M_{\rm t}$ implies, the direction of
the net spin density around one impurity site is parallel with that
around the other impurity site. This is easily understood as a manifestation
of the dominant AF tendency of the impurity-free
half-filled honeycomb
lattice. The net majority moments around both impurities are mainly on 
sublattice B, and the nearest-neighbor AF interaction implies a ferromagnetic
interaction between spins on the same sublattice.
We show the dependence of the moment at each lattice point on the distance
from impurity sites in Fig. \ref{fig:fig4} (a).
It can indeed be seen that the moment is localized around impurities
and ferrimagnetic short-range order develops.

\begin{figure}[hbt]
\begin{center}
\includegraphics[width=6cm]{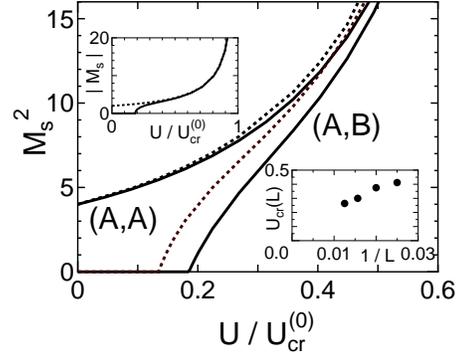}
\caption{
The dependence of the square of staggered moment $M_{\rm s}$ on the interaction
strength $U$; the linear size of the lattice is $L=40$ (solid curves) and
$L=64$ (dotted curves). 
$M_{\rm s}$ varies smoothly in the case (A,A) where both
impurities are on sublattice A, while
it evolves from zero at an $L$-dependent critical $U_{\rm cr}(L)$
in the case (A,B) where one impurity is on sublattice A and the other is
on sublattice B. The upper inset shows the dependence of $M_{\rm s}$ 
on $U$ in a wider range for $L=40$. 
The lower inset shows the dependence of $U_{\rm cr}(L)$ on $1/L$.
The critical value $U_{\rm cr}^{(0)}$ for the bulk honeycomb lattice
is $U_{\rm cr}^{(0)}=2.23 t$.
}
\label{fig:Ms_dep}
\end{center}
\end{figure}

\begin{figure}[hbt]
\begin{center}
\includegraphics[width=6cm]{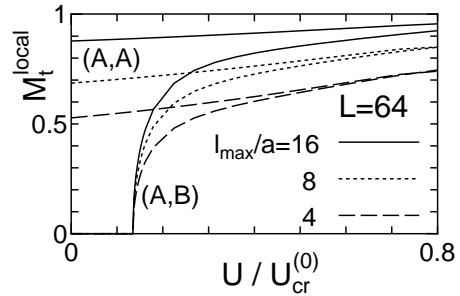}
\caption{
Local spin moment $M_{\rm t}^{\ell}(\ell_{\rm max})$ as a function of
$U$ for $L=40$: $\ell_{\rm max}=16a$ (solid curves), $8a$ (dotted curves),
and $4a$ (dashed curves).
}
\label{fig:M_local}
\end{center}
\end{figure}

The dependence of the square of staggered moment $M_{\rm s}=m_{A}-
m_{B}$ on $U$
is shown in Fig. \ref{fig:Ms_dep}; $M_{s}$ smoothly changes from
2 as $U$ increases from zero.
Total sublattice moments $m_{A}$ and $m_{B}$ are given by
$m_{A}=(2+M_{\rm s})/2$ and $m_{B}=(2-M_{\rm s})/2$ ($M_{\rm s}<0$).
This means that no moment is induced on sublattice A as $U\rightarrow
0$. This is because the zero-mode wave function vanishes on sublattice
A.\cite{kuma2,Waka2} The sublattice moment on sublattice A is induced
by a finite $U$. This can be also seen in Figs. \ref{fig:spind} (b)
and \ref{fig:fig4} (a).
The local moment $M_{\rm t}^{\ell}(\ell_{\rm max})$, defined by
$M_{\rm t}^{\ell}(\ell_{\rm max})={\sum_{\mib{n}}}'(m_{\mib{n}A}-
m_{\mib{n}B})$ where the summation is restricted to those lattice points
within the distance of $\ell_{\rm max}a$ from an impurity site,
is also shown as a function of $U$ in Fig. \ref{fig:M_local}.
In contrast to the staggered moment $M_{\rm s}$, the local moment
varies only modestly.\cite{unity} 
It is the 
staggered moment that is enhanced by the interaction.

Next, we study the case where
one vacancy is on sublattice A, and the other is on sublattice B; we
denote this case by (A,B).
In this case, the zero modes appear pairwise above and below the
chemical potential (see Fig. \ref{fig:AA_diagr} (b)).
At any $U$, two of the zero modes are occupied by up- and down-spins
and the total moment $M_{\rm t}$ always vanishes
in harmony with Lieb's theorem.\cite{Lieb}
At $U=0$, the occupied zero-mode wave function $\phi_{0\sigma}^{(-)}(\bfr)$
takes a finite value only on sublattice $\alpha$ around an impurity site 
on sublattice $\overline{\alpha}$.\cite{kuma2,small}
The spin density at each site
vanishes, because 
$\phi_{0\uparrow}^{(-)}(\bfr)=\phi_{0\downarrow}^{(-)}(\bfr)$.
As $U$ increases, this identity is violated. For example,
$|\phi_{0\uparrow}^{(-)}(\bfr)|$ takes a larger value on sublattice B
than $|\phi_{0\downarrow}^{(-)}(\bfr)|$, and vice versa on sublattice A.
This results in a net up spin density around the impurity site
on sublattice A and a net down spin density around the impurity
site on sublattice B.
The spin density around an impurity site at $U=1.2 t$
is very similar to the one in the case (A,A) at the same value of $U$, 
{\it i.e.},  Fig. \ref{fig:spind} (a). 
We again find 
short-range ferrimagnetic order around a vacancy site.
See also Fig. \ref{fig:fig4} (b).
In this case, the induced net moment around the impurity site
on sublattice A is antiparallel to that around the impurity site
on sublattice B. This is also understood as a result of the
inherent AF tendency of the half-filled honeycomb lattice.
Actually, the net moment induced around one impurity is 
equal in magnitude with that induced around the other impurity.

We find that, for a finite $L$, the linear size of the lattice, there is
a critical value of $U$, $U_{\rm cr}(L)$, above which staggered moment 
takes a finite value, as can be seen from Fig. \ref{fig:Ms_dep}.
Near $U_{\rm cr}(L)$, the dependence of $M_{\rm s}$ 
is well approximated as $M_{\rm s}\propto \sqrt{U-U_{\rm cr}(L)}$.
The dependence of $U_{\rm cr}(L)$ on $L$ is shown in one of the insets of
Fig. \ref{fig:Ms_dep}. The previous
results\cite{kuma1,kuma2} strongly suggested that the staggered 
susceptibility on the non-interacting half-filled honeycomb lattice
diverges in the presence of a strong impurity potential. This implies
that $U_{\rm cr}(L)$ vanishes as $L\rightarrow\infty$, that is,
a finite staggered moment is induced by an infinitesimally small value
of $U$. Moreover, as can be seen from Fig. \ref{fig:Ms_dep},
the $|M_{\rm s}|-U$ curve in the (A,B) case merges that in the
(A,A) case as $U$ increases, and it merges at a smaller $U$ at a larger
$L$. This also implies that $U_{\rm cr}(L)$ vanishes as $L\rightarrow
\infty$. Indeed, it will make no difference whether two impurities
are on the same sublattice or on the different sublattices,
if the distance between them goes to infinity. From these arguments,\cite{deg}
it is clear that $U_{\rm cr}(L\rightarrow\infty)$ vanishes, although
we cannot definitely 
confirm $U_{\rm cr}(L\rightarrow\infty)=0$ from the
present calculation
because of the limited size of the used lattice.
The dependence of the local moment $M_{\rm t}^{\ell}(\ell_{\rm max})$
on $U$ is shown in Fig. \ref{fig:M_local}. It also merges 
$M_{\rm t}^{\ell}(\ell_{\rm max})$ in the (A,A) case as $U$ increases.

The local moment formation is closely related to the presence of
the zero modes, {\it i.e.}, it is specific to the half-filled honeycomb
lattice. By doing similar calculations for an off-half-filled honeycomb
lattice and a half-filled square lattice (without nesting),
we indeed find that no local moment formation occurs at $U$ much smaller than
$U_{\rm cr}^{(0)}$. Vacancies on the same sublattice points in a row
generate a zigzag edge (and a bearded edge) where ferrimagnetc
order is realized.\cite{Fuji,Oka} This order at a zigzag edge is thus
closely connected to the short-range order around a vacancy studied here.

\begin{figure}[hbt]
\begin{center}
\includegraphics[width=5cm]{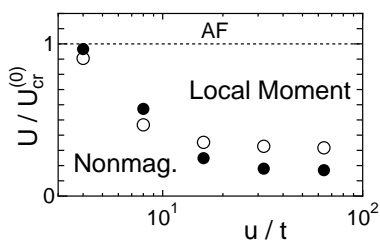}
\caption{
Region where local moments are formed on the $u-U$ plane in the (A,B) case
obtained for $L=20$ (open dots) and $50$ (filled dots). Electron number
$N_{\rm e}$ is equal to $N_{\rm L}-N_{\rm i}=N_{\rm L}-2$.
At $U>U_{\rm cr}^{(0)}$, the AF state is realized.
}
\label{fig:Uu}
\end{center}
\end{figure}
Finally, we briefly discuss the case with finite $u$. In Fig. \ref{fig:Uu},
we show the region where a local moment is formed around an impurity
site in the case (A,B). 
At $u \grtsim 10 t$, results converge to those at $u\rightarrow\infty$.
At smaller $u$, the region where a local moment is formed is rather
limited. Finite size effects are clearly seen, and the detailed
study of them is left for a future study.

Now, we discuss the applicability of the present result to actual
graphene. We do not know the precise value of $U$ appropriate
for graphene, and it is also difficult to estimate the value of $u$ that 
represents the effect of a defect. Moreover, we have neglected the
effect of long-range part of the Coulomb interaction and the possible
local distortion around a defect.
Yazyev and Helm recently found ferrimagnetic short-range
order around a vacancy and also around a hydrogen chemisorption 
defect using a first-principles calculation.\cite{yaz}
This implies that actual graphene with nonmagnetic defects is in the
region of Fig. \ref{fig:Uu} where local moments develop around
impurities and that the short-range order is caused by the same
mechanism discussed in this work.

We find that the energy gain per vacancy due to the moment formation
is approximately $0.01 t$,\cite{gain} {\it i.e.}, 300 K in graphene.
This implies that
it is difficult to observe local moment formation around a single vacancy
at room temperatures, and that it can be observed only at low temperatures.
It is possible that a moment develops more stably
near a few defects close to each other. This is also a
subject of a future study.

A local magnetic moment can also be formed around a defect
in other carbon-based materials. For example, oscillation of spin density
was found around a vacancy in a (10,0) carbon nanotube with a 
first-principles calculation.\cite{ma2} 
This may also be caused by the same mechanism discussed here.

To conclude, we have studied the effect of nonmagnetic defects on magnetism
of electrons on the half-filled honeycomb lattice and found that
short-range ferrimagnetic order is induced and a local moment is
formed around a defect. This strongly implies 
that a nonmagnetic defect in graphene
indeed induces a local magnetic moment around it.

We would like to thank Dr. K. Wakabayashi and Dr. O. V. Yazyev
for useful discussions and correspondence. This work was
supported in part by Research Foundation for the Electrotechnology
of Chubu.

\appendix


\begin{thebibliography}{99}
\bibitem{move}
A chemisorbed hydrogen atom is mobile, but we do not consider the mobility
here.
\bibitem{charge}
Charge density is found to be uniform except for the impurity sites, although
the spin density spatially varies.
\bibitem{unity}
If we sum the moments in the half of the total lattice points,
{\it i.e.}, in the rectangle with side-lengths $L/2$ and $\sqrt{3}L/2$
around an impurity, the local moment is unity for any $U$.
\bibitem{small} The zero-mode wave function on sublattice $\alpha$
around an impurity site on sublattice $\alpha$ does not vanish
exactly, but is negligibly small and tends to vanish when 
the distance between the two impurities goes to infinity.
\bibitem{deg} At $L\rightarrow \infty$, the four zero modes becomes degenerate
at $U=0$ as in the (A,A) case, and the local moments will be induced by 
an infinitesimally small $U$.
\bibitem{gain} The energy gain per vacancy is $0.0104 t$ at $U=1.0 t$ for
$L=64$ in the (A,B) case.
\end{thebibliography}
\end{document}